\documentclass[english,aps,pre, reprint, twocolumn]{revtex4}
\usepackage[T1]{fontenc}
\usepackage[latin9]{inputenc}
\usepackage{float}
\usepackage{color}
\usepackage{amsmath}
\usepackage{graphicx}
\usepackage[caption=false]{subfig}
\makeatletter
\@ifundefined{textcolor}{}
{%
 \definecolor{BLACK}{gray}{0}
 \definecolor{WHITE}{gray}{1}
 \definecolor{RED}{rgb}{1,0,0}
 \definecolor{GREEN}{rgb}{0,1,0}
 \definecolor{BLUE}{rgb}{0,0,1}
 \definecolor{CYAN}{cmyk}{1,0,0,0}
 \definecolor{MAGENTA}{cmyk}{0,1,0,0}
 \definecolor{YELLOW}{cmyk}{0,0,1,0}
}

\makeatother

\usepackage{babel}
\begin{document}
\title{Self-assembly and complex manipulation of colloidal mesoscopic particles by active thermocapillary stress}

\author{Subhrokoli Ghosh}
\author{Aritra Biswas}
\affiliation{Dept of Physical Sciences, Indian Institute of Science Education
and Research Kolkata, Mohanpur, Kolkata, India 741235}
\author{Basudev Roy}
\affiliation{Dept of Physics, Indian Institute of Technology Madras, India 600036}
\author{Ayan Banerjee}
\email{ayan@iiserkol.ac.in}
\affiliation{Dept of Physical Sciences, Indian Institute of Science Education
and Research Kolkata, Mohanpur, Kolkata, India 741235}
\begin{abstract}
We demonstrate that the active thermocapillary stresses induced by multiple microbubbles offer simple routes to directed self-assembly and complex but controllable micromanipulation of mesoscopic colloidal particles embedded in a liquid. The microbubbles are nucleated on a liquid-glass interface using optical tweezers. The flow around a single bubble causes self-assembly of the particles in rings at the bubble-base, while an asymmetric temperature profile generated across the bubble interface breaks the azimuthal symmetry of the flow, and induces simultaneous accumulation and repulsion of particles at different axial planes with respect to the bubble. The flows due to two adjacent bubbles leads to more diverse effects including the sorting of particles, and to local vorticity that causes radial and axial rotation of the particles - the latter being obtained for the first time using optical tweezers. The sorting is enabled by nucleating the bubbles on spatially discrete temperature profiles, while the vorticity is generated by nucleating them in the presence of a temperature gradient which once again causes a strong symmetry-breaking in the azimuthal flow. The flow profiles obtained in the experiments are explained by analytical solutions or qualitative explanations of the associated thermocapillary problem employing the Stokes and heat equations.
\end{abstract}
\maketitle

%\rmfamily 
\section*{Introduction}

The study of active stress on mesoscopic particles in fluidic environments
has evoked widespread scientific interest in recent times since it
enables the understanding and ultimate control of fundamental processes
involving fluid-particle interactions at a microscopic level, and
leads to a variety of applications. The fundamental processes range
from transport of organelles inside the fluidic environment in a cell
\citep{leal2007advanced}, to the motion of cilia and flagella of
swimming bacteria \citep{kim2005}. The analysis of the active stress
involved in many such processes has led to the design of ``active''
particles \citep{brennen1977,softmatterrevire2013}, that often mimic
natural microswimmers \citep{programmablemicroswimmer,singh2015many}.
These active particles move in pre-designed paths by interacting with
their environment and exchanging energy by diverse processes including
thermophoresis \citep{thermophoresis1980,thermophoresis2010,di2009colloidal},
electrophoresis \citep{electro1993capillary}, diffusiophoresis \citep{diffusiphoresis2012},
etc. Such controlled motion of mesoscopic particles under different
active stress has very significant implications in different areas
including cell biology \citep{cellmobility}, nanomedicine \citep{nanomedicine},
and micro-patterning \citep{cellmicro2007}. Active stress can also
be induced on microparticles in a fluid by manipulating the flow environment
using different techniques of actuation \citep{actuator2011} in microfluidic
channels, and by generating surface stresses by controlled modulation
of fluid-fluid or fluid-solid boundaries \citep{surfacestress2005,stres1999}.
The latter can be achieved by a number of processes such as electrowetting
\citep{electrowetting}, thermocapillary action \citep{thermocapillary,thermobubble},
and Marangoni stresses \citep{pearson1958convection,marangoni_hawaii}.

Recently, it has been shown that the thermocapillary flow around a
thermo-optically generated bubble can be used to manipulate colloidal particles \citep{zhao2014,lin2018,zheng2011} cells
\citep{bubblecell}, various nano-particles \citep{armon2017,namura2016sheathless,namura2016,zno2017}, quantum dots \citep{bangalore2017,lin2015} and carbon
nanotubes \citep{mathurbubble}. These bubbles are generated by focusing
the trapping laser on an absorbing substrate on the wall of a water
filled trapping chamber \citep{hagalangmuir}. The rapid heating causes
the water in the vicinity of the laser ``hot spot'' to spinodally
decompose and grow into a bubble. The bubble reaches a steady state
size determined by the laser power and remains thermophoretically
trapped in the region of the temperature maximum. Since there is an
appreciable temperature difference between the pole of the bubble
which is in direct contact with the hot spot and its opposite pole
which is farthest from the hot spot, an active Marangoni stress is
immediately developed around the bubble surface. This active stress
drives fluid flow towards the bubble and away from the wall. Particles
are drawn towards the bubble due to this convective flow and attach
to the bubble-wall interface \citep{mathurbubble,broylangmuir}. The
hot spot can be translated and the bubble, driven by the strong thermophoretic
forces, follows it. This provides a method for transporting particles
that are attached to the bubble surface.

Such thermophoretic trapping mechanisms have several advantages of
over standard optical tweezers. First, thermophoretic traps are agnostic
to the dielectric contrast between the particle and the suspending
medium. Thus, particles with negative dielectric constant that are
impossible to trap using standard optical means can easily be trapped
thermophoretically. Second, thermophoretic traps can apply nN forces
while optical traps can only apply much weaker forces in the range
of a few hundred pN. Third, thermophoretic traps can transport multiple
particles over large distances stably and simultaneously \citep{bubblemicrorobot}.
While such trapping and manipulation by attaching particles to the
bubble surface has been experimentally demonstrated \citep{bubblecell,bubblemicrorobot,mathurbubble},
non-contact manipulation of particle trajectories by generating diverse
flow patterns by bubbles is missing. Understandably, a clear picture
of the active stress driven hydrodynamic flow is also lacking. In
this paper, we address both these issues. Experimentally, we demonstrate
interesting and reproducible self assembly of polystyrene microparticles
on a microbubble surface, and proceed to engineer novel flow patterns
in the vicinity of a single and a pair of microbubbles, which induce
interesting motion in polystyrene microparticles that trace the flow
lines. Thus, we demonstrate simultaneous attraction and repulsion
of microparticles in different axial planes around a bubble by inducing
a temperature asymmetry along the azimuthal direction. Next, we grow
bubbles adjacent to each other in two different configurations - first
with the bubbles separated and having non-overlapping temperature
profiles around them, and second with the two bubbles grown along
a temperature gradient. The former leads to distinct sorting of microparticles
depending upon the size of the adjacent bubbles with more particles
accumulating on the larger bubble, while the latter produces vorticity
in the radial and axial directions, such that microparticles sampling
the flow lines exhibit rotational motion in three dimensions, akin to
that produced in optical traps designed with angular momentum (both
longitudinal and transverse - the latter being notoriously difficult
to achieve for propagating light \cite{photonicwheels})-carrying
laser beams. By solving the Stokes and heat equations with appropriate
boundary conditions, we estimate the flow pattern of the fluid around
a laser induced microbubble corresponding to different experimental
situations, and obtain excellent qualitative agreement. We now proceed
to report our experimental observations.

\section*{Experimental results\label{sec:expres-1}}

The thermo-optic tweezers setup is based on an inverted Olympus IX71
microscope with a 100x 1.3 NA objective lens focusing the tweezers
laser beam at 1064 nm into the sample chamber containing an aqueous
dispersion of the sample (usually polystyrene microspheres of different
diameters). We use a surface that is pre-coated by linear patterns
of a Mb-based soft oxometalate (SOM) material by a method developed
by us \citep{broylangmuir,ghoshjmcc}. The SOM material has finite
absorption at 1064 nm, so that a hot spot is formed leading to a bubble
when the tweezers laser is focused on any region along the pattern.
The absorptive surfaces (typically a standard glass coverslip which
forms the bottom surface of the chamber or a microscope slide which
forms the top surface of the chamber) are attached to each other by
double sided sticky tape. In both cases, the temperature of the hot
spot would vary between 373K (the threshold for bubble formation)
to 644K - the latter corresponding to the critical temperature for
water after which it cannot exist as a liquid with the surface tension
tending to zero \citep{njcbasudev}. The temperature of the hot spot
can be modified by changing the laser intensity incident on it, which
we typically achieve by changing the laser power. For some of the
experiments, we couple another laser of the same wavelength into the
system. This is done in three ways: 1) Focus the two lasers on a single
pattern with the separation so small, that the two separate bubbles
which are formed merge to form a single bubble. Now, the patterned
SOM is partially conducting, with a conductivity of around 50 S/cm
\citep{ghoshjmcc}. Thus, we have a single bubble with a temperature
asymmetry in the vicinity of the hot spot in the azimuthal direction.
2) Focus the lasers on a single pattern with the beam spots well separated,
so that separate stable bubbles are formed. This changes the aysmmetry
of the azimuthal temperature profile compared to 1). 3) Focus the
two lasers on two separate spatially isolated linear patterns. Thus,
two different bubbles are nucleated with similar temperature profiles
around the hot spot. These lead to entirely different flows that can
be traced using the polystyrene microspheres freely diffusing in the
sample. 

\begin{figure}[h]
	\includegraphics[width=0.5\textwidth]{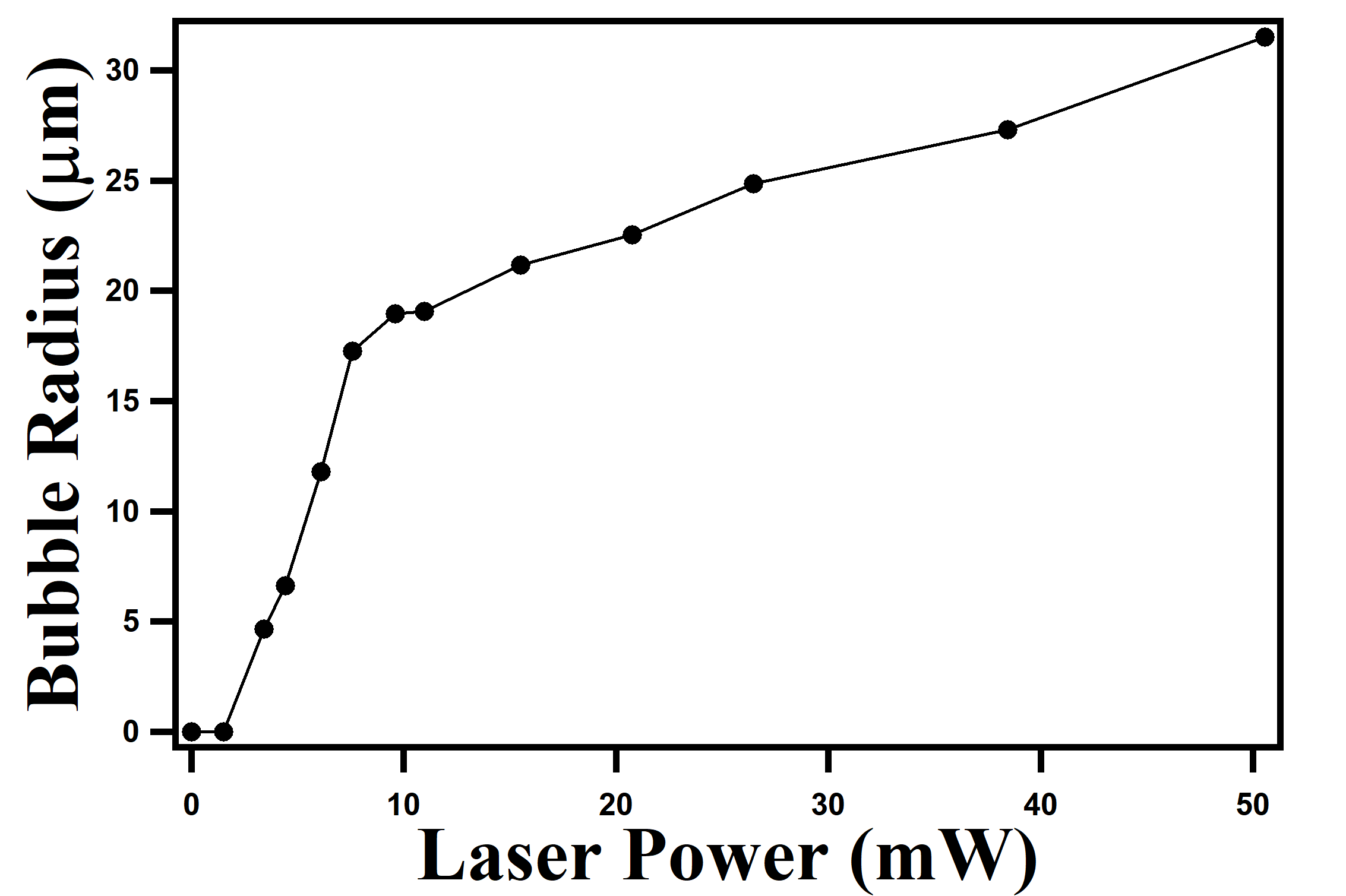}\caption{\label{fig:RvsP}Bubble radius as a function of laser power incident on the sample.}
\end{figure}

Before going into the detailed experimental results for the different
configurations, we first confirm the fact that the heat transfer happening
at the bubble surface is diffusive in nature and not convective. To
confirm this, we perform a simple experiment, where we measure the
size of a bubble as a function of laser power\citep{namura2017}. Thus, the tweezers
laser beam is incident on a precoated SOM pattern in the sample chamber
to generate a bubble. After the bubble is generated, the radius of
the bubble is measured after sufficient amount of time (10 seconds
for instance) so that the bubble size reaches a quasi-steady state
where it remains constant over a few minutes. Then the laser is turned
off and the bubble shrinks subsequently. The same experiment is repeated
with different laser powers and the result is shown in figure \ref{fig:RvsP}.
From the graph \ref{fig:RvsP} we can conclude that,

\begin{enumerate}
	\item the bubble formation starts at a finite laser power, which is of the
	order of 3 mW in this case.
	\item The bubble grows rapidly at first up to a certain radius, after which the rate of growth slows down significantly.
\end{enumerate}

After several such runs, we determine that the limiting size of the
bubble up to which it grows rapidly is around$\sim20$ $\mu m$ .
Now, in an earlier work \citep{njcbasudev}, we have shown that this
fast growth corresponds to heat transfer from the hot spot through
the liquid by diffusion, but the growth is arrested at the onset of
convection, when heat is dissipated much more efficiently by the liquid.
This can be understandable from the simple fact that for small bubbles
the temperature difference between the opposite poles (the point of
the bubble in contact with the hot spot and the point vertically opposite
point to it) is correspondingly small, so that convection is not observed.
The situation changes when the bubble size increases, which increases
the temperature difference so that convection switches on after a
certain temperature difference is achieved. Thus, when the laser power
is low, diffusion dominates the bubble growth, while after a threshold
power (corresponding to a bubble radius of $\sim20$ $\mu m$), convection
takes over. This is a crucial finding, which sets the limiting size
of the bubbles in our subsequent experiments, as we can only assume
the Stokes flow to be valid when the energy equation obeys Laplace
equation, ie, heat transfer is dominated by diffusion. We shall revisit
this formulation again in Section \ref{sec:Theory}). We now proceed
to the experimental results for different configurations.

\subsection*{Self assembly of particles on bubbles\label{subsec:selfassem-1}}
\begin{figure}
	\centering\includegraphics[width=0.48\textwidth]{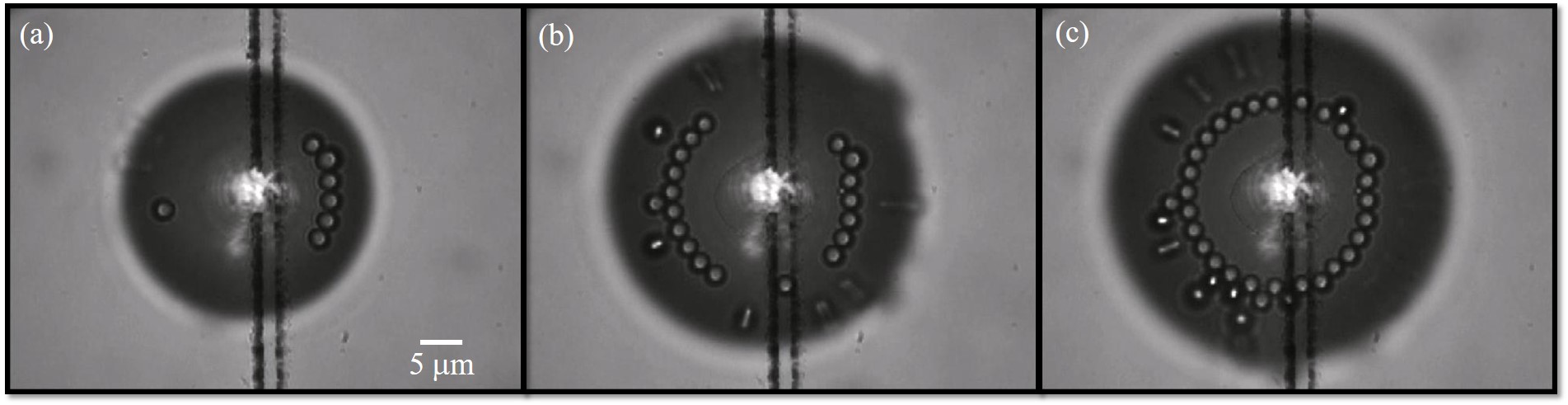}
	
	\caption{\label{fig1} Time lapse images of 3 $\mu\mathrm{m}$ diameter polystyrene microparticles in aqueous dispersion assembling on a bubble of initial diameter around 30 $\mu\mathrm{m}$ . The microparticles follow flow	lines and assemble around the base of the bubble. We see more particles in (b) panel a which is at t= 25s, while a complete	ring forms in (c) with the bubble size arouund 40 $\mu\mathrm{m}$,	at t=35s. The images have been extracted from Video 1 in the multimedia files of the Supplementary Information. }
\end{figure}

\begin{figure} [H]
	\includegraphics[width=0.48\textwidth]{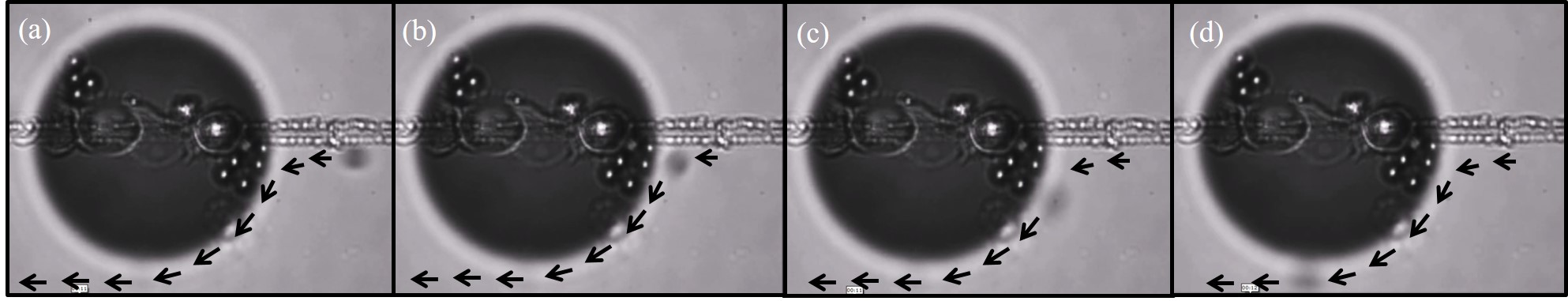}\caption{\label{fig:solidangle}Schematic diagram to determine the contact angle subtended at the base of a bubble.}
\end{figure}

This experiment is performed using a single trapping laser coupled
into the sample chamber and focused on the linear SOM pattern to nucleate
a bubble. Once the bubble is formed, convective flows set in which
draw particles towards it. As we demonstrate in the theoretical analysis
which follows, the flow lines tend to converge towards inward direction
to the bubble, so that we have an assembly of particles near that
region. The particles are attracted towards the bubble and finally
settle at the bubble- wall interface, where they form a ring around
the bubble base. One can determine the size of the ring in the method demonstrated in Fig.~\ref{fig1} from a simple estimation of the contact angle shown in Fig.~\ref{fig:solidangle}. Note that particles of different diameters form rings of different sizes for a fixed contact angle i.e. for the same size of the bubble. One such ring is shown in Fig.\ref{fig1}, where we have a bubble of initial diameter around 30 $\mu\mathrm{m}$ formed along
a SOM trail seen behind the bubble. Polystyrene microspheres of diameter
3 $\mu\mathrm{m}$ follow the flow lines and assemble with time as
shown in Fig.\ref{fig1}a - c. Video 1 in the multimedia files of
the Supplementary Information shows the entire process in real time.
We observe self assembly of a variety of particles including polystyrene
beads of diameter 1, 3, and 10 $\mu\mathrm{m}$ (Video 2), as well
as other particles including streptavidin coated magnetic beads of
diameter 3 $\mu\mathrm{m}$ (Video 3), and gold micro particles (Video
4a) and silver nanoparticles (Video 4b) of around 100-1000 nm in size.
Note that the magnetic and metallic nanoparticles are notoriously
difficult to trap using conventional optical tweezers due to their
large scattering cross-sections. However, using this technique the
scattering is rendered irrelevant since the particles are not exposed
to light at all. Besides self-assembly around the bubble, all such types of particles can be controllably transported by the same bubble.

\subsection*{Particle sorting using single and two bubbles \label{partsort-1}}

\subsubsection*{Sorting using a single bubble:}

Here we employ Method 1 mentioned in Section \ref{sec:expres-1},
where two laser beams of wavelength 1064 nm and equal power are focused very close to each other on the same SOM pattern. This gives rise to a single bubble (we observe nucleation of two bubbles initially,which soon merge into a single bubble) with a temperature profile that varies azimuthally.

\begin{figure}[h]
	\centering
	\includegraphics[width=0.48\textwidth]{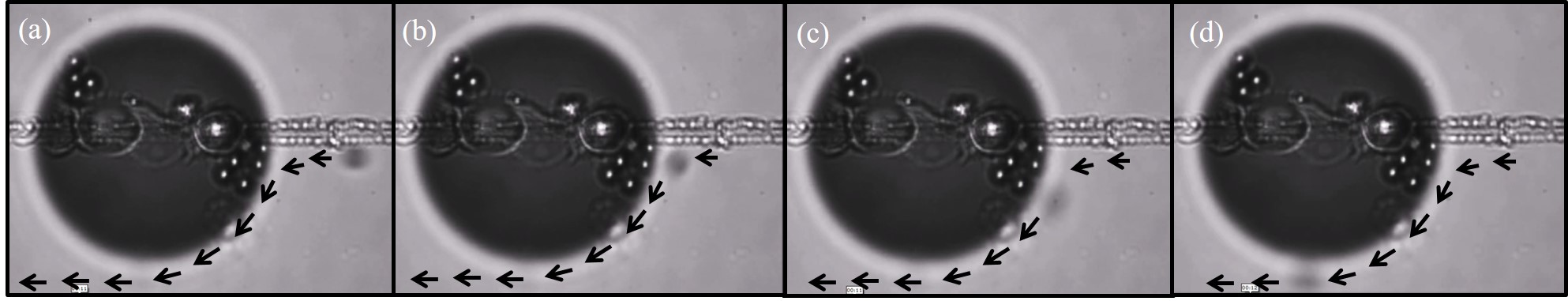}
	\caption{Simultaneous accumulation and pumping at different axial positions
		using a single bubble with a temperature gradient (created by two
		laser beams focused on the bubble surface) across it. Microparticles
		on the same focal plane as the image accumulate on the surface of
		the bubble where the flow streamlines converge, whereas those at a
		different axial plane (shown here in the form of an unfocused microparticle)
		are pumped away. The trajectory of the unfocused particle has been
		depicted using black arrows, and the images occur at progressively
		later times. The images have been extracted from Video 5 in the multimedia
		files of the Supplementary Information. \label{fig2}}
\end{figure}

The flow generated by this design leads to the situation depicted in Figs. \ref{fig2} (a) - (d), which are again time lapsed images of Video 5 in the SI. Thus, we have a set of particles which accumulate on the surface of the bubble-wall interface, and are observed in sharp focus in the figure, while other particles on different z-planes are repelled away. In the figure, we demonstrate this using a single tracer particle which appears unfocused. Thus,the particle approaches the bubble in Figs. \ref{fig2} (a) and (b),while it is deflected away in Figs. \ref{fig2} (c) and (d). The trajectory of the particle is qualitatively represented by the use of black arrows.This phenomenon is also captured in Fig.\ref{fig1theory}, where the tracer particle follows the trajectories in X-Z plane. It should therefore be clear that by changing the size of the bubble, it should be able to sort particles which are diffusing about in a particular plane in the sample. Modifying the temperature asymmetry should also affect the z-plane(s) where particles are repelled. This can enable spectroscopy of mesoscopic particles without requiring them to be subjected to the high intensities associated with optical tweezers.

\subsubsection*{Sorting using two bubbles:}

\begin{figure}
	\includegraphics[width=0.48\textwidth]{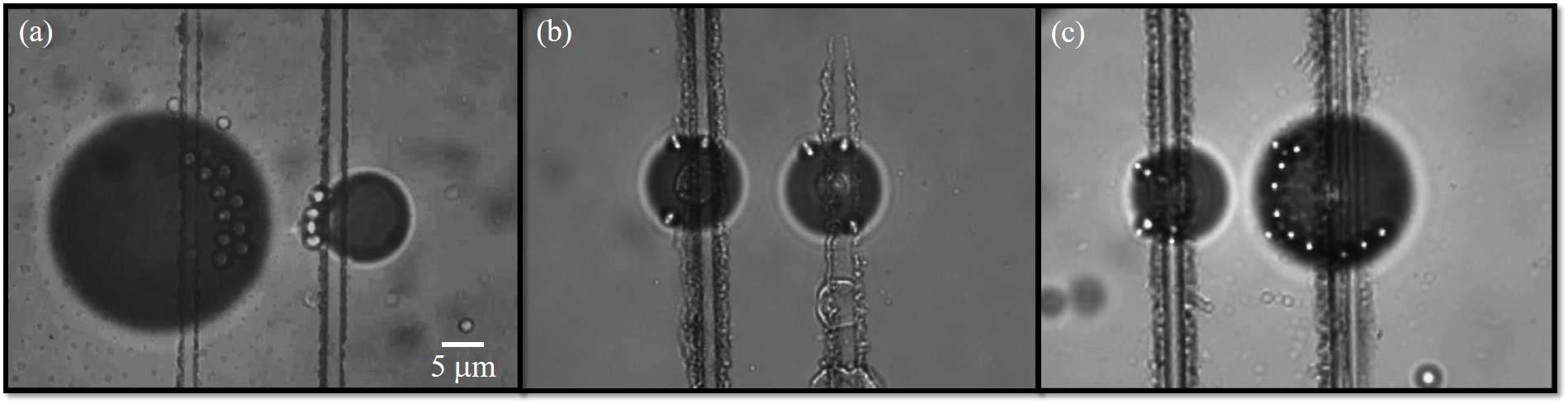}
	
	\caption{Sorting polystyrene particles using a pair of bubbles. (a) Left bubble
		bigger than the right, as a result of which more particles are observed
		to be attached on the former. (b) Bubbles are of the same size - each
		bubble has four particles attached. (3) Opposite case of (a) - more
		particles adhere to the bubble on the right now.\label{fig3}}
\end{figure}

In these set of experiments, we created two adjacent bubbles on neighbouring
pre-existing SOM patterns. These are shown in Figs. \ref{fig3}(a)-(c)
(adapted from Videos 6-8 in the SI), where we have different size
ratios of the bubbles in each case, namely 2.3:1, 1:1, 1:1.75, going
from left to right. The effects of the resultant flow patterns are
apparent from the assembly of the tracer particles assembled on the
bubbles. It is clear that a separatrix can be identified in the flow
between the two bubbles, and particles would be drawn towards a bubble
depending on whether its trajectory lies above or below the separatrix.
The position of the separatrix is modified with the change in size
of the bubbles, and is more towards the smaller bubble as is clear
from the figures, where we observe an equal number of particles assembled
in Fig. \ref{fig3}(b) where the bubble sizes are equal. We can thus
envisage the use of this technique to generally sort particles by
size modulation of the two bubbles, where the position of the separatrix
would be continuously modulated by changing the laser power, so that
by a suitable choice of separation between the bubbles and modulation
frequency, one can allow particles of a certain diameter to pass between
the bubbles while others would be trapped on the bubble surfaces \citep{ultrasound_sorting}.
We are currently working on some of these experiments.

\subsection*{Inducing angular momentum in particles using vortex flows \label{partmanipulate-1}}

Inducing orbital motion in mesoscopic particles in optical tweezers
typically involves the use of angular momentum carrying optical beams,
which are non-trivial to generate. In addition, creating multiple
traps with opposite angular momentum increases the complexity of the
required experimental apparatus, and yet have interesting significance
in a variety of experiments involving hydrodynamic coupling \citep{grier2006anomalous,grier2003revolution,dholakia_angular},
synchronization \citep{di2012hydrodynamic}, etc. To the best of our
knowledge, generating angular momentum in mesoscopic particles without
the use of angular momentum carrying beams has not yet been demonstrated.

\begin{figure}
	\centering\includegraphics[width=0.48\textwidth]{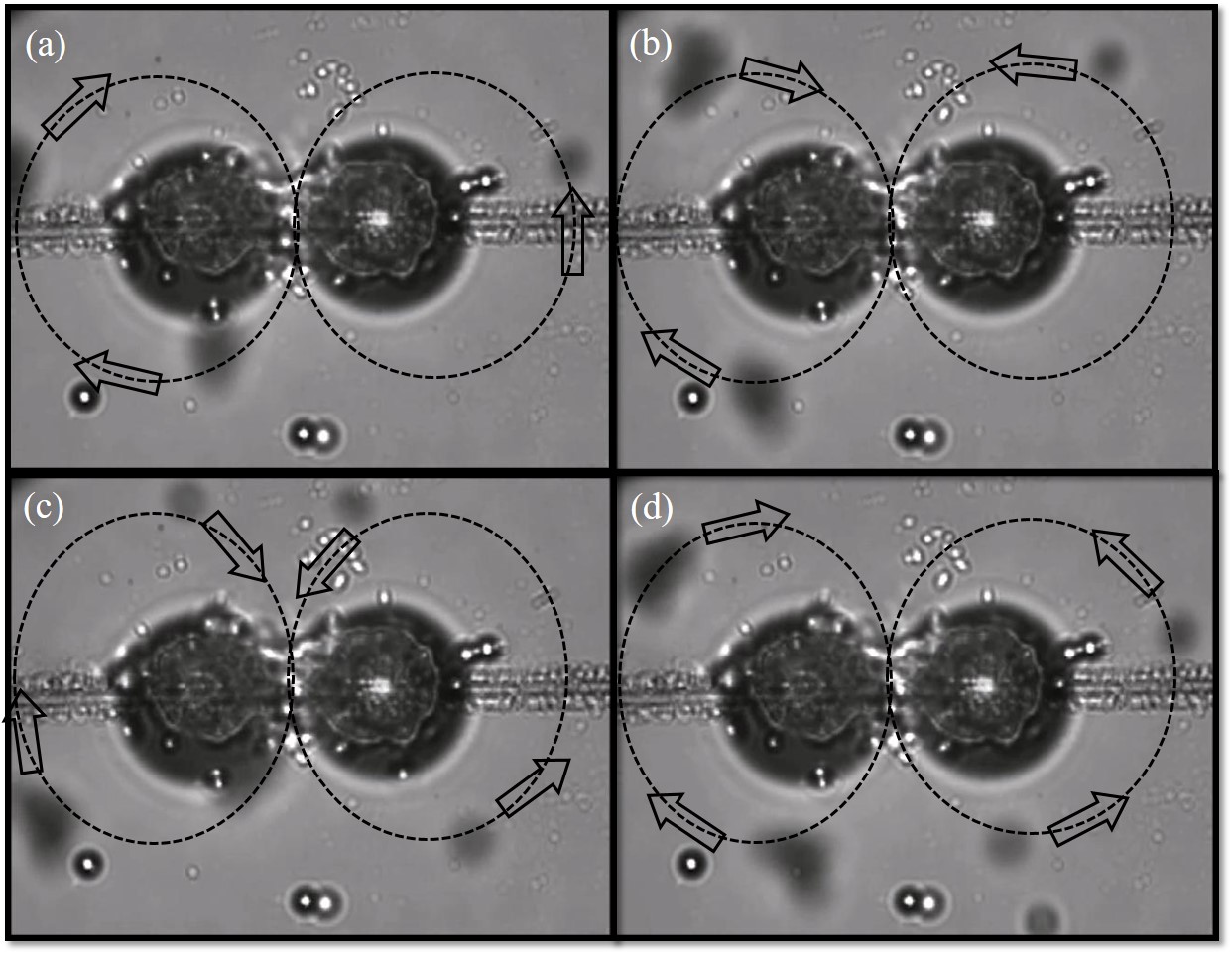}
	\caption{Time lapsed images showing the circular trajectory of polystyrene
		microparticles in the vicinity of two bubbles grown along the same
		SOM pattern. The particles - which appear unfocused - are at different
		axial planes with respect to the focal plane of the image, and undergo
		opposite circular trajectories around each bubble. Images (a) - (d)
		are at progressively later time. The images have been extracted from
		Video 9 in the multimedia files of the Supplementary Information.
		\label{fig4}}
\end{figure}

\begin{figure}
	\centering\includegraphics[width=0.48\textwidth]{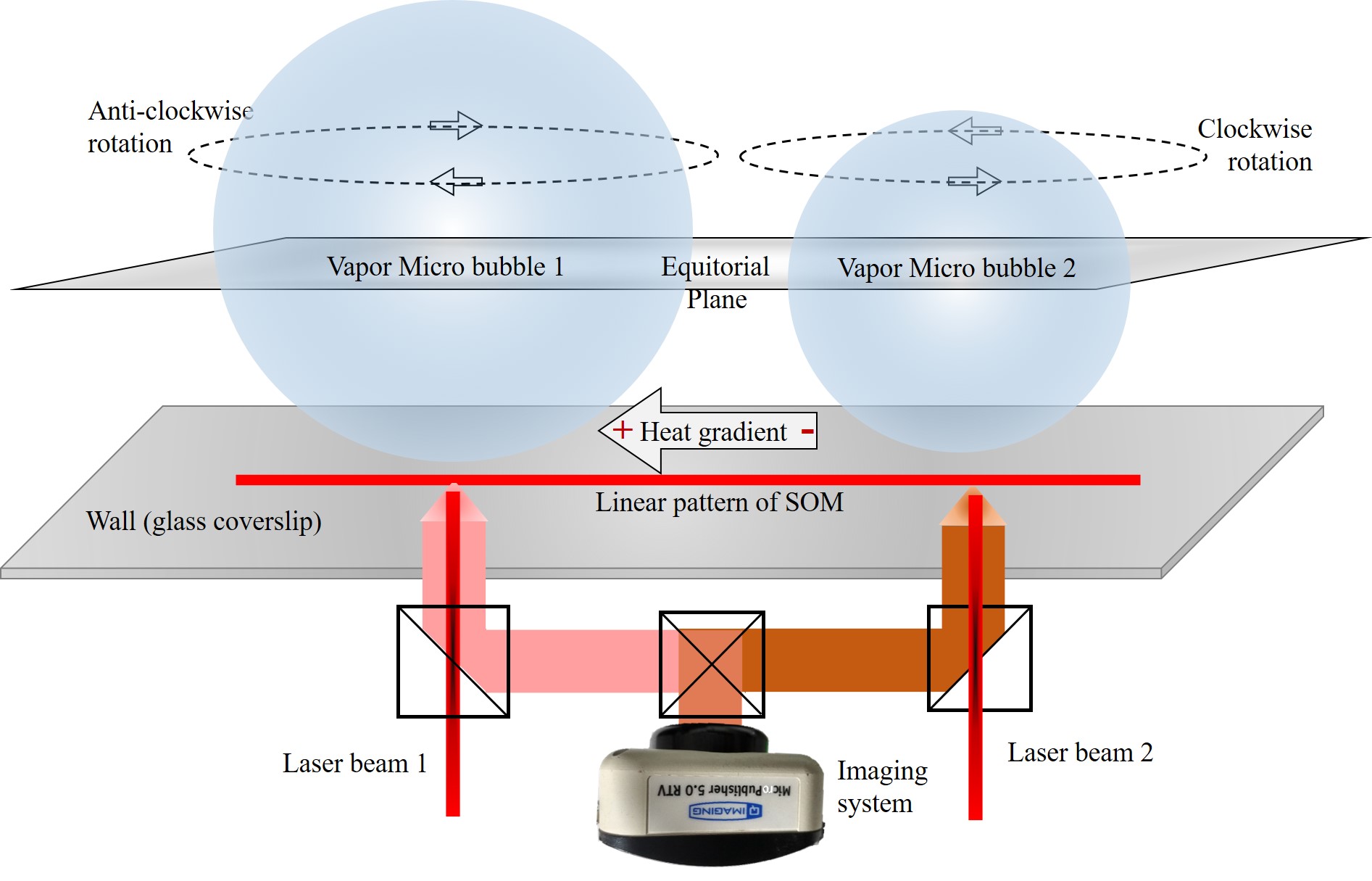}
	\caption{Schematic diagram of the sense of rotation of probe particles around the bubbles of unequal sizes}
	\label{rotation_schematic}
\end{figure}

\begin{figure}[h]
	\includegraphics[width=0.48\textwidth]{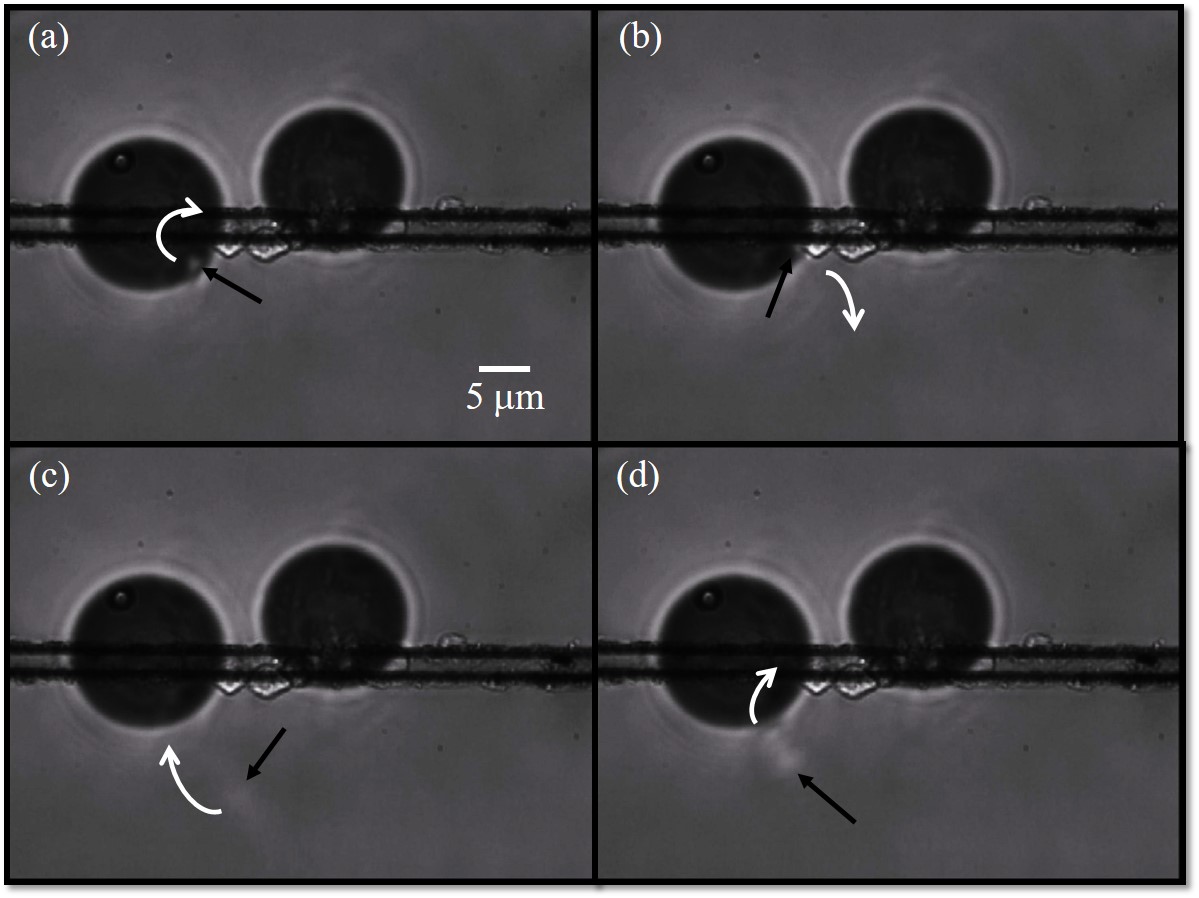}
	\caption{Axial rotation of the probe signifying three dimensional vortices
		around the two bubbles grown along the same SOM pattern. (a)-(d) show
		time lapsed images taken from Video 13. The probe is in good focus
		in (a), but since it is moving in the axial direction, gradually goes
		out of focus in each subsequent image until it is barely seen in (c),
		but is partially in focus in (d) again. The black arrows show the
		location of the probe, while the white arrows depict its trajectory.
		\label{fig6}}
\end{figure}

Here, we demonstrate this by modifying our experimental design in
the manner mentioned in Method (3) of Sec. \ref{sec:expres-1}. Thus,
we grow two bubbles on the same SOM pattern with their centers separated
far enough so that the bubbles do not merge.  The laser power in each beam is the same, as is apparent from the size of the bubbles demonstrated in Fig. \ref{fig4}. Interestingly, we now observe tracer particles
exhibiting orbital motion in opposite directions at different axial
planes with respect to the bubbles. The full video is available in
Video 9 in the SI, here we provide time lapsed images once again.
The trajectory of the tracer particles in the vicinity of the bubbles
is shown by dotted lines. The instantaneous positions of unfocused
particles executing the flows is shown in Figs. \ref{fig4}(a)-(d).
Note that we observe particles randomly entering the flow, and also
leaving it - possibly due to local perturbations. Interestingly, we
observe clockwise and anti-clockwise flows near the left and right
bubble, respectively. Now, the direction of the flow is typically dependent on the size of the bubble - for similar sized bubbles, we hypothesise that the flow direction is governed by a spontaneous symmetry breaking condition driven by the temperature gradient in the proximity of the bubble. Thus, both clockwise and anti-clockwise flows are possible in this condition around each bubble, and we explicitly demonstrate this again in Video 10 in the SI, where we have two similar sized bubbles but with anti-clockwise flow around the left bubble. We discuss this condition in greater detail in the next section. When we have bubbles whose diameters are very different, we typically observe anti-clockwise flow around the bubble which is bigger, and clockwise around that 	which is smaller. This has been demonstrated in Video 11 in the	SI, where the right bubble is much smaller than the left, and we observe clockwise flow around it. The opposite case happens when we have the right bubble much bigger than the left one, so that a particle rotates anti-clockwise as we demonstrate in Video 12. Note that these are examples of explicit symmetry breaking in the flow profile, where we always obtain the same flow direction given a particular configuration of bubbles. Figure \ref{rotation_schematic} depicts this situation schematically. We are also able to observe axial rotation in the probes as is clear in Videos 13 and 14, time lapsed images from which we show in Fig. \ref{fig6}(a)-(d). Here, we show 	the location of the probe using black arrows while the axial trajectory	is depicted by white arrows. It is clear that the polystyrene probe gradually goes out of focus between (a)-(c), is barely visible in	(c), before coming back near focus in (d) (the motion is much better	observed in Video 13). This implies that its $z-$position is being altered with respect to the microscope imaging plane. This is due to the presence of three-dimensional flows in the system, the axial component of which is predominantly sampled by the probe. Note that even for the axial flow we obtain both clockwise and anti-clockwise	rotation as is depicted in Videos 13 and 14, respectively. Such axial rotation of a probe particle is extremely difficult to achieve in optical tweezers, and is generally by optical beams referred to as ``photonic wheels'' in the literature \citep{photonicwheels}. Photonic wheels are a figurative description of the presence of transverse spin angular momentum in the light - which can lead to axial rotation of particles, and has been mostly reported for evanescent fields \citep{photonicwheels}.

The active stresses that we can manufacture using our methods can
thus directly facilitate such complex three dimensional rotation in
optical micromanipulation. The angular momentum induced in the flow
- as shown in Videos 9-14 in the SI - is facilitated by the temperature
gradient on the linear pattern formed by the presence of two heat
sources (and therefore two bubbles). This is confirmed from Video
15 in the SI, where we show that the axial rotation of the probe gradually
comes to a stop when one of the heat sources is withdrawn (by turning
off the laser), so that the corresponding bubble shrinks, and the
temperature gradient is thereby destroyed. 

We now develop a theoretical formalism to explain the experimental
observations.

\section*{Theory\label{sec:Theory}}

We consider the following assumptions to be true for our system:
\begin{enumerate}
	\item The effects of gravitation is neglected.
	\item Inside the bubble, the thermal conductivity and viscosity of the gas
	is considered to be negligible compared to that of the fluid outside.
	\item The bubble remains spherical throughout, and deformations are assumed
	to be negligible.
\end{enumerate}
So the generalised continuity, energy and momentum equations of our
system can be written as, \begin{subequations}
	\begin{gather}
	\frac{\partial\rho}{\partial t}+\nabla.(\rho\textbf{v})=0\label{eq:continuity}\\
	\frac{\partial\theta}{\partial t}-\nabla[-\textbf{v}\theta+\alpha\nabla\theta]=0\label{eq:energy}\\
	\rho\frac{\partial\textbf{v}}{\partial t}+\nabla[\textbf{v}.\textbf{v}-\eta(\nabla\textbf{v}+\nabla\textbf{v}^{T})+\textbf{p}]=0\label{eq:momentum}
	\end{gather}
\end{subequations}
Here $\rho$ is the density of the fluid, $\alpha$
is the thermal diffusivity, $\eta$ is the coefficient of dynamic
viscosity, $\theta$ is the temperature, p is the external pressure
and $v$ represents the fluid flow . In the incomprecible limit, equation
\ref{eq:continuity} becomes 
\begin{gather*}
\boldsymbol{\nabla}\cdot\boldsymbol{v}=0.
\end{gather*}
The P\'{e}clet number $(P_{e})$, which is the ratio of the rate of advective
transport to the rate of diffusive transport for the heat transfer
is given as, 

\begin{gather*}
P_{e}=\frac{v\theta}{\alpha(\nabla\theta)}\sim\frac{vL}{\alpha}\sim10^{-5}
\end{gather*}
as for our system the characteristic length scale $L$ is of the order
of $10^{-6}m$, typical velocity is of the order of $10^{-6}m/s$
and the thermal diffusivity of water is of the order of $10^{-7}m^{2}/s$
respectively. So, we can neglect the advective term compared to the
diffusive term in the equation \ref{eq:energy}. Again, The Reynolds
number $(R)$, defined as the ratio of the inertial to the viscous
forces, is given by 
\begin{gather*}
R=\frac{\rho vL}{\eta}=\frac{vL}{\nu}\sim10^{-6}
\end{gather*}
where, the kinematic viscosity ($\nu$) of water at room temperature
is $\sim10^{-6}~Pa.s$. So, the inertial term can be neglected compared
to the diffusive term in the equation \ref{eq:momentum}.
\begin{figure}[t]
	\includegraphics[clip,width=0.4\textwidth]{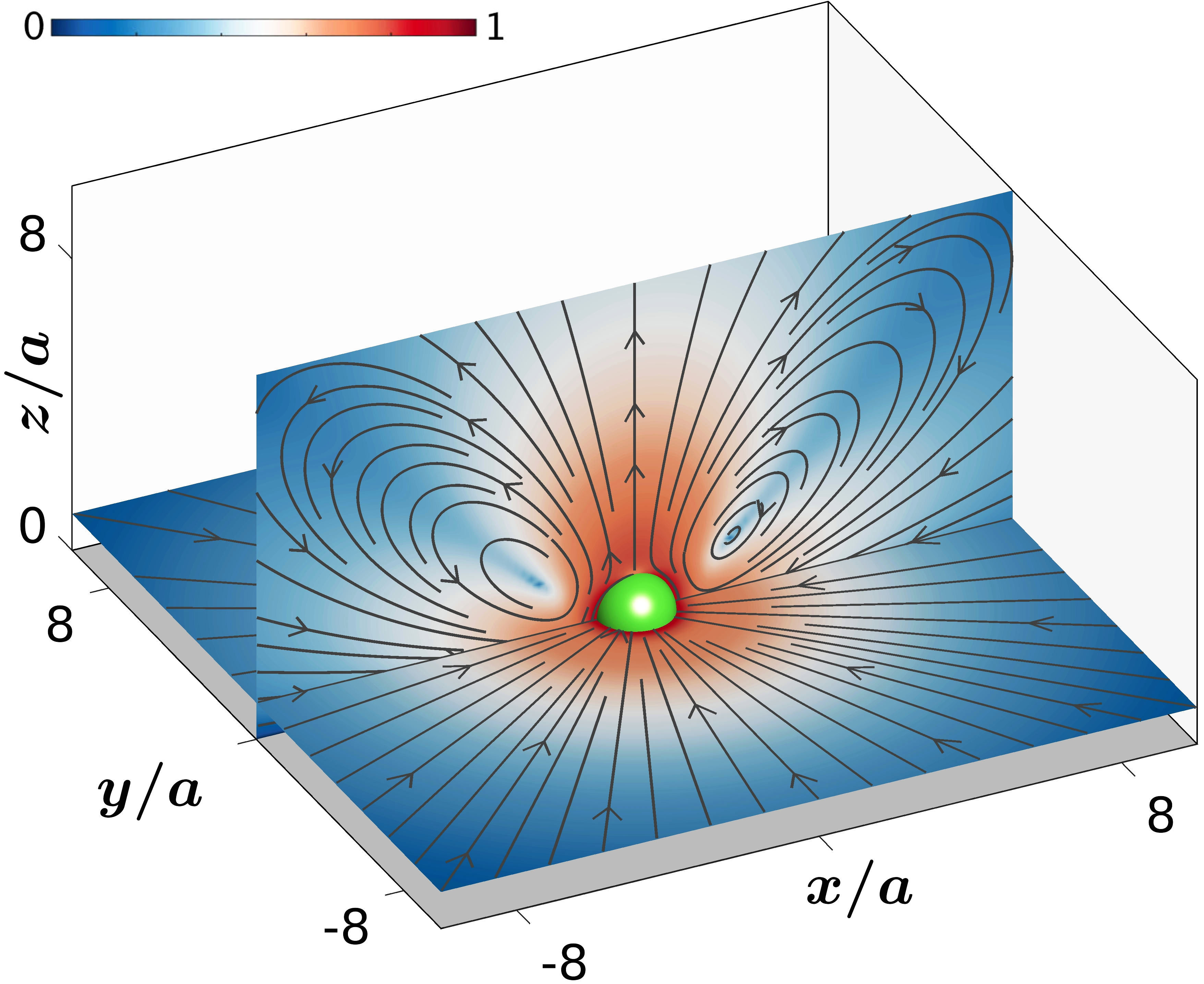}
	\caption{\label{fig1theory} Cross-section of the streamlines of fluid flow
		around a bubble in planes parallel and perpendicular to the wall (shown
		in grey at $z=0$). It can be seen that the flow has cylindrical symmery
		and draws fluid from all direction. This explains the aggregation
		of tracer particles on the surface of bubbles in Fig.(\ref{fig1})
		and Video 1 in SI. This also explains the simultaneous accumulation
		and pumping of the tracer particles at different axial planes of the
		bubble in Fig.(\ref{fig2}) and Video 5 in the SI. In addition, as
		explicit from the figure itself, the axial rotation demonstrated in
		Fig.(\ref{fig6}) and Video 12 of SI, is also captured by this. The
		streamlines of the fluid flow are drawn over the pseudo-color plot
		of the normalized logarithm of the flow speed.}
\end{figure}

\begin{figure}
	\centering\includegraphics[width=0.5\textwidth]{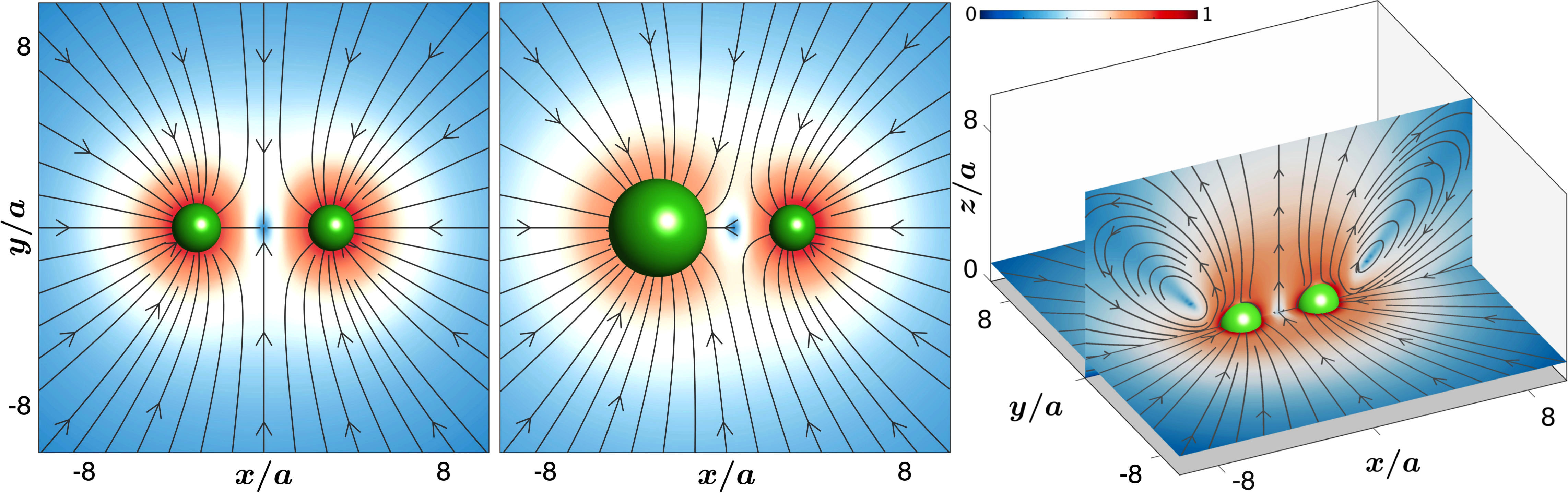}\caption{Fluid flow around the two bubbles for the first experimental case, where there is
		only one bubble per pattern. These flow lines are obtained from the
		linear superposition of the flow around of a single bubble in Fig.(\ref{fig1theory}).
		Left: Fluid flow around two bubbles of same size. A separatrix is
		formed at the symmetric plane between the two bubble. Middle: Fluid
		flow around a bigger and a smaller bubble. In this case the flow is
		predominantly towards the larger bubble. Right: Three-dimensional
		representation of the fluid flow around two bubbles of same size.
		The streamlines are overlaid on the the pseudo-color plot of the normalized
		logarithm of the flow speed.\label{fig:2theory}}
\end{figure}

From dimensional analysis, the ratio of the advective to the diffusive
term for the equation \ref{eq:momentum} of our system is 
\begin{gather*}
\frac{v.v}{\eta(\nabla v)}=\frac{vL}{\eta}\sim10^{-6}
\end{gather*}
which is very small. So, the advective term can also be ignored as
well in the equation \ref{eq:momentum} compared to the diffusive
term. Again, in equilibrium, all the time dependent terms in equation
$1$ go to zero. So, considering all these, let us assume that we have
a spherical air bubble of radius $a$ in an incompressible fluid of
viscosity $\eta$ bounded by a plane wall at $z=0$\textit{.} The
center of the bubble is at $\boldsymbol{R}$, while $\boldsymbol{r}$
denotes a point on the surface of the bubble. The fluid flow$v(\boldsymbol{r})$,
at a point $\boldsymbol{r}$ in the fluid, satisfies the flowing equations
as, 
\begin{subequations}
	\begin{gather}
	\bf{\nabla}\cdot\bf{v}=0.\\
	\nabla^{2}\theta=0.\label{eq:heat}\\
	\bf{\nabla}\cdot\bf{\sigma}=0.\label{eq:stokes}
	\end{gather}
\end{subequations}
where$\bf{\sigma}=-p\bf{I}+ \eta (\nabla \bf{v} + (\nabla \bf{v})^{T})$ 
is the Cauchy stress and $p$ is the fluid pressure \citep{happel1965low},
equation \ref{eq:stokes} is the Stokes equation of fluid flow. We
assume there is no motion normal to the interface and hence a vanishing
normal velocity there. The normal Laplace pressure is balanced by
the interfacial tension and the tangential Marangoni stress is balanced by the viscous stress. These lead to the following boundary conditions on the surface $S$ of the bubble

\begin{subequations}\label{eq:fluidBC}
	\begin{align}
	\bf{v}_{n} & =\hat{\rho} \hat{\rho}\cdot\bf{v}=0 \\
	\bf{f}_{t} & =(\bf{I}-\hat{\rho}\hat{\rho})\cdot\bf{f}=-\Gamma'\bf{\nabla}_{s}\theta,\label{eq:tangBalance}
	\end{align}
\end{subequations}
where $\boldsymbol{f}=\boldsymbol{\hat{\rho}}\cdot\boldsymbol{\sigma}$
is the traction on the interface and $\boldsymbol{\rho}$ is the radius
vector from the center of the bubble to a point on its surface. The
temperature field $\theta$ and its tangential gradient $\boldsymbol{\nabla}_{s}\theta$
are assumed to be known, while $\Gamma'=\partial\Gamma/\partial\theta$,
where $\Gamma$ is the surface tension. In addition, the fluid flow
vanishes at the plane wall and at at very large distances from it,
ie, \begin{subequations}
	\begin{align}
	\boldsymbol{v}=0\text{ at}\,\,\,\, & z=0,\\
	\boldsymbol{v}\rightarrow0,\,\,\,\, & r\rightarrow\infty,
	\end{align}
\end{subequations}

The temperature field $\theta(\boldsymbol{r})$ satisfies the steady
heat equation (\ref{eq:heat}) with a no flux boundary condition on
the bubble interface and a relaxation to the imposed temperature field
at infinity:\begin{subequations}
	\begin{align}
	\hat{\boldsymbol{\rho}}\cdot\boldsymbol{\nabla}\theta=0,\label{eq:gradTemp}\\
	\theta\rightarrow\theta^{\infty},\text{ as } & r\rightarrow\infty.
	\end{align}
\end{subequations} The vanishing of the heat flux is consistent with
the negligible heat capacity of air that in the interior of the bubble.
The imposed temperature field also satisfies the steady-state heat
equation $\nabla^{2}\theta^{\infty}=0.$ As the energy equation is
decoupled from the momentum equation, due to the assumption of negligible
advective transport of energy, the temperature field can be solved
for independently and the solution inserted to obtain the Marangoni
stress needed to solve the velocity equation.

\subsection*{Integral equation for temperature }

We use the boundary integral method to obtain the solution of the
temperature field \citep{leal2007advanced}. In this approach, the
temperature at a point $\mathbf{r}$ in the bulk is obtained from
the boundary integral representation of the Laplace equation, which
is given as 
\begin{align}
\theta(\mathbf{r}) & =\theta^{{\scriptscriptstyle \infty}}(\mathbf{r})-\int\Phi(\mathbf{r},\,\mathbf{R}+\mathbf{\boldsymbol{\rho}})\,\hat{\boldsymbol{\rho}}\cdot\boldsymbol{\nabla}\theta(\mathbf{r})\,\text{d}S.\nonumber \\
& +\int\hat{\boldsymbol{\rho}}\cdot\mathbf{\boldsymbol{\nabla}}\Phi(\mathbf{r},\,\mathbf{R}+\mathbf{\boldsymbol{\rho}})\,\theta(\mathbf{r})\,\text{d}S.\label{eq:bieLaplace}
\end{align}
Here $\theta^{{\scriptscriptstyle \infty}}(\mathbf{r})$ is the externally
imposed temperature and $\Phi(\mathbf{r},\,\mathbf{r}')$ is a Green's
function of Laplace equation
\begin{alignat*}{1}
\nabla^{2}\Phi(\mathbf{r},\,\mathbf{r}')=-\delta(\mathbf{r}-\mathbf{r}').
\end{alignat*}
Evaluating the boundary integral representation on the boundary of
the bubble, accounting for the singular nature of the second integral,
and imposing the boundary condition (see Eq.(\ref{eq:gradTemp}))
on the first integral gives the following integral equation for the
temperature distribution on the interface
\begin{align}
\frac{1}{2}\theta(\mathbf{r}) & =\theta^{{\scriptscriptstyle \infty}}(\mathbf{r})+\int\hat{\boldsymbol{\rho}}\cdot\mathbf{\boldsymbol{\nabla}}\Phi(\mathbf{r},\,\mathbf{R}+\mathbf{\boldsymbol{\rho}})\,\theta(\mathbf{r})\,\text{d}S.\label{eq:bieTemp}
\end{align}
This is a Fredholm integral equation of the second kind with a symbolic
solution
\begin{equation}
\theta=(\frac{1}{2}-\mathcal{K})^{-1}\theta^{\infty},
\end{equation}
where $\mathcal{K}$ is the double layer integral operator. 
We now obtain the explicit form of the above formal solution using the Galerkin
method by expanding the temperature fields in tensorial spherical
harmonics $\mathbf{Y}^{(l)}$ as
\begin{equation}
\theta(\bf{r})=\frac{(2l+1)}{4\pi}\sum_{l=0}^{\infty}\bf{\Theta}^{(l)}\cdot\bf{Y}^{(l)}(\hat{\rho}).\label{eq:tempExpansion}
\end{equation}
The unknown expansion coefficients $\bf{\Theta}^{(l)}$ are
then determined by the integral equation. The tensorial spherical
harmonics are defined as 
\begin{equation}
\bf{Y}^{(l)}(\hat{\rho})=(-1)^{l}\rho^{l+1}\nabla^{(l)}\rho^{-1}.\label{harmonics}
\end{equation}
The tensorial spherical harmonics form a orthogonal set of basis functions
on the surface of a sphere
\begin{alignat}{1}
\frac{1}{4\pi b^{2}}\int\bf{Y}^{(l)}(\hat{\rho})\,\mathbf{Y}^{(l')}(\hat{\rho})d\text{S} & =\delta_{ll'}\,\frac{l!\,(2l-1)!!}{(2l+1)}\mathbf{\Delta}^{(l)}.\label{eq:orthogonality}
\end{alignat}
Here $\mathbf{\Delta}^{(l)}$ is tensor of rank $2l$, projecting
any \textit{l}th order tensor to its symmetric irreducible form. 

Inserting the tensorial expansion in the integral representation for
the temperature, Eq.(\ref{eq:bieLaplace}) and expanding the Green's
function about the origin of the sphere yields
\begin{align}
\theta(\boldsymbol{r})=\theta^{{\scriptscriptstyle \infty}}(\boldsymbol{r})+\sum_{l=1}^{\infty}b^{l+1}\boldsymbol{\nabla}^{(l)}\Phi\,\cdot\boldsymbol{\Theta}^{(l)}.
\end{align}
which gives the temperature field at any point in the bulk in terms
of the Green's function and the expansion coefficients. Thus, exploiting
the spherical shape, we obtain exact analytical expressions for the
boundary integrals in terms of a Green's function \citep{singh2018generalized}.

\begin{figure}[!t]
	\centering\includegraphics[width=0.5\textwidth]{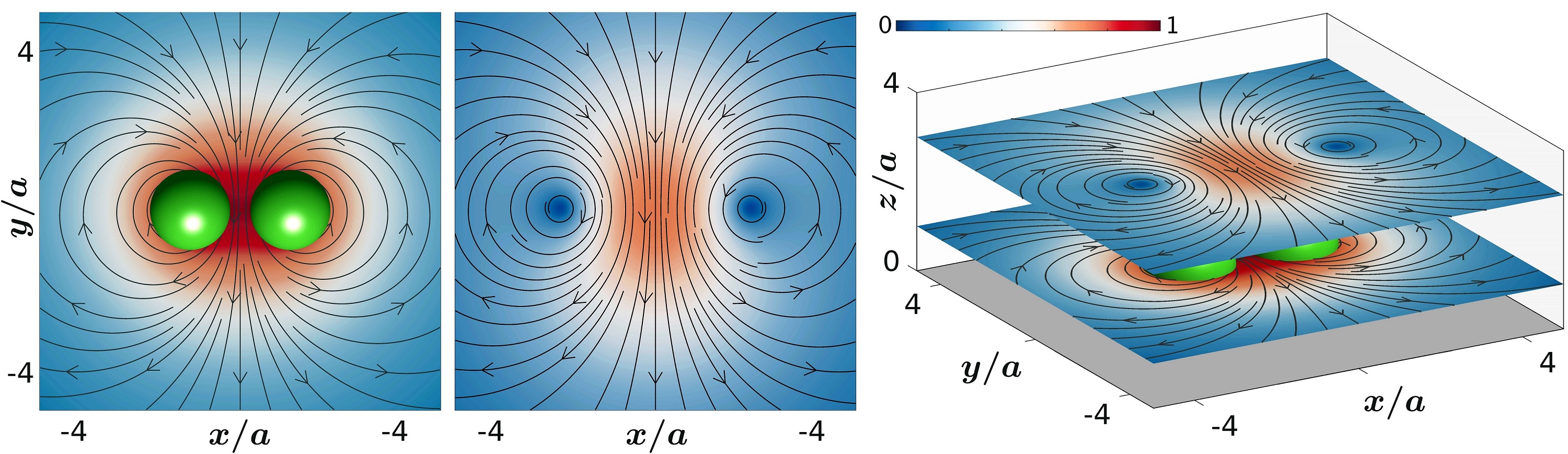}
	\caption{Fluid flow around the two bubbles on the same linear pattern. Left
		panel shows the flow above the equatorial plane of the bubble, at
		$z=3a$. The middle panel shows the flow in the equatorial plane of
		the bubble, ie, $z=b$, while the wall (shown in grey) is at $z=0$.
		The right panel shows a three-dimensional plot of the streamlines
		around two bubbles.\label{fig:3theory}}
\end{figure}

We now consider the boundary integral equation for the temperature
on the surface of the bubble. We multiply both sides of this integral
by the \textit{l}-th basis function and integrate over the surface.
Tensorial harmonics are eigenfunctions of the double layer which results
in a diagonal system of equations whose solutions is 
\begin{align}
\boldsymbol{\Theta}^{(l)} & =\frac{4\pi b^{l}}{l!(2l+1)!!}\boldsymbol{\nabla}^{(l)}\theta^{{\scriptscriptstyle \infty}}(\boldsymbol{R}).
\end{align}
The temperature distribution on the interface is then 
\begin{alignat}{1}
\theta(\boldsymbol{r})= & \,\theta^{{\scriptscriptstyle \infty}}+b\boldsymbol{\nabla}\theta^{\infty}\cdot\mathbf{Y}^{(1)}+\frac{b^{2}}{6}\boldsymbol{\nabla}\boldsymbol{\nabla}\theta^{\infty}\cdot\mathbf{Y}^{(2)}\nonumber \\
+\, & {\frac{b^{3}}{90}}\boldsymbol{\nabla}\boldsymbol{\nabla}\boldsymbol{\nabla}\theta^{\infty}\cdot\mathbf{Y}^{(3)}+O(\boldsymbol{\nabla}\boldsymbol{\nabla}\boldsymbol{\nabla}\boldsymbol{\nabla}\theta^{\infty}).\label{eq:tempBubbleSurface}
\end{alignat}
Here, the derivative of the imposed distribution are evaluated at
the center $\boldsymbol{R}$ of the bubble. 

The externally imposed temperature $\theta^{\infty}$ leads to the
surface tension $\Gamma$, which is assumed to be linearly dependent
on the temperature field $\theta(\boldsymbol{r})$ of the bubble surface.
The tangential derivative, $\boldsymbol{\nabla}_{s}=(\boldsymbol{I}-\hat{\rho}\hat{\rho})\cdot\boldsymbol{\nabla}$,
of the temperature field, then provides the Marangoni stress in the
boundary condition (Eq.(\ref{eq:tangBalance})) for the flow field.
Thus, the force on the bubble is obtained in terms of the surface
integral of the tangential gradient of the surface temperature \citep{subramanian1985}.
Explicitly, it is given as
\begin{align}
\mathbf{F}^{\text{c}} & =-\tfrac{1}{2}\Gamma'\int\boldsymbol{\nabla}_{s}\theta\,dS,\label{eq:forceM}\\
\mathbf{T}^{\text{c}} & =-\Gamma'\int\boldsymbol{\rho}\times\boldsymbol{\nabla}_{s}\theta\,dS.\label{eq:torqueM}
\end{align}
Using Eq.(\ref{eq:tempBubbleSurface}) in the above expressions gives
forces and torques in terms of the external imposed temperature field
$\theta^{\infty}$. From Eq.(\ref{eq:forceM}), it is clear that an
average gradient of externally imposed field leads to a force on the
bubble. In the experiments, the bubble is stationary at the wall,
and thus these Marangoni force has to be counterbalanced by force
on the bubble. It is precisely these forces which determine the exterior
fluid flow and determine the dynamics of tracer particles. In the
next section, we use the above form to obtain the exterior fluid flow
around the bubbles and discuss the dynamics observed in the experiments.

\subsection*{Exterior fluid flow}
We use the singularity method for Stokes flow to obtain the exterior
flow around the bubbles near a plane wall \citep{pozrikidis1992}.
The singularity method of solution is constructed out of the fundamental
solutions of the Stokes equation. These fundamental solutions are
the Green\textquoteright s functions of the Stokes flow. We use the
Lorentz-Blake tensor, which satisfies the no-slip condition at a plane
wall. 
The expression for the fluid flow due to a single bubble in
singularity method, due to a force $\mathbf{F}$ and a torque $\mathbf{T}$,
is given as
\begin{equation}
\boldsymbol{v}(\boldsymbol{r})=\mathbf{G}^{\text{w}}\cdot\mathbf{F}+\frac{1}{2}\nabla\times\mathbf{G}^{\text{w}}\cdot\mathbf{T}.\label{eq:flow}
\end{equation}
Here $\mathbf{G}^{\text{w}}$ is the Lorentz-Blake tensor \citep{blake1971c},
whose explicit form is given as
\begin{align}
G_{\alpha\beta}^{\text{w}}(\boldsymbol{r},\,\boldsymbol{R}) & =\tfrac{1}{8\pi\eta}\left[G_{\alpha\beta}^{0}(\boldsymbol{r}-\boldsymbol{R})+G_{\alpha\beta}^{*}(\boldsymbol{r},\,\boldsymbol{R}^{*})\right],\label{eq:lorentzBlake}
\end{align}
where
\[
\boldsymbol{G}^{0}(\boldsymbol{r})=(\nabla^{2}\boldsymbol{I}-\boldsymbol{\nabla}\boldsymbol{\nabla})\,r,
\]
is the Oseen tensor, and $\boldsymbol{G}^{*}(\boldsymbol{r},\,\boldsymbol{R}^{*})$
is the contribution from the image system to satisfy the boundary
condition at the no-slip wall. The contribution from the image system
is \citep{blake1971c}
\begin{gather}
G_{\alpha\beta}^{*}(\boldsymbol{r}^{*})=-\frac{\delta_{\alpha\beta}}{r^{*}}-\frac{r_{\alpha}^{*}r_{\beta}^{*}}{r^{*^{3}}}+2h^{2}\bigg(\frac{\delta_{\alpha\nu}}{r^{*^{3}}}-\frac{3r_{\alpha}r_{\nu}}{r^{*^{5}}}\bigg)\mathcal{M}_{\nu\beta}\nonumber \\
-2h\bigg(\frac{r_{3}^{*}\delta_{\alpha\nu}+\delta_{\nu3}r_{\alpha}^{*}-\delta_{\alpha3}r_{\nu}^{*}}{r^{*3}}-\frac{3r_{\alpha}^{*}r_{\nu}^{*}r_{3}^{*}}{r^{*^{5}}}\bigg)\mathcal{M}_{\nu\beta}.
\end{gather}
Here $\boldsymbol{R}^{*}=\boldsymbol{\mathcal{M}}\cdot\boldsymbol{R}$,
while $\boldsymbol{\mathcal{M}}=\boldsymbol{I}-2\mathbf{\hat{z}}\mathbf{\hat{z}}$
is the mirror operator, while \textit{h} is the height of the center of the bubble from the wall. Since the bubble is stationary and attached to the wall ($h=R$),
the force $\mathbf{F}=-\mathbf{F}^{\text{c}}$ and the torque $\mathbf{T}=-\mathbf{T}^{\text{c}}$are
obtained from the counterbalance of the Marangoni force in Eq.(\ref{eq:forceM}).
We compute the flow due to two bubbles, in the superposition limit,
by adding the contributions of Eq.(\ref{eq:flow}). The above analysis
is then used to plot the fluid flow around the bubbles due to cases
(a) and (b) in Fig.(\ref{fig:2theory}) and Fig.(\ref{fig:3theory})
respectively and compare them with the experiments.

We classify the experimental system in two classes for simplicity
- (i) one bubble on a pattern or two bubbles on two separate patterns,
and (ii) two bubbles on a pattern. We now consider each of these cases
separately below.

For the first case, the external imposed temperature field on the
bubble can be approximated as 
\begin{equation}
\theta^{\infty}(\boldsymbol{r})=c_{{\scriptscriptstyle 0}}-c_{1}\cos\vartheta,\label{eq:caseA}
\end{equation}
where $c_{{\scriptscriptstyle 0}}$ and $c_{1}$ are constants and
$\vartheta$ is the polar angle of the spherical polar coordinate
($\boldsymbol{\hat{\rho}},\boldsymbol{\hat{\vartheta}},\boldsymbol{\hat{\varphi}}$).
This is then used to compute the magnitude of the force on the bubble
using Eq. (\ref{eq:forceM}). The torque vanishes, since the temperature
distribution is axisymmetric, while there is a constant force acting
in the direction perpendicular to the wall. The streamlines of the
fluid flow around a bubble is plotted in Fig.(\ref{fig1theory}).
It can be seen that there is an inward flow toward the bubble in the
plane of the wall. This explains the aggregation of the tracer particles
on the bubble surface in the experiments, as shown in Fig.(\ref{fig1})
and Video 1 in SI. The simultaneous accumulation and pumping of the
tracer particles at different axial planes of the bubble in Fig.(\ref{fig2})
and Video 5 in the SI can also be mapped from the Fig.(\ref{fig1theory}).
Clearly, the in-plane accumulation in Fig.(\ref{fig1}) and Fig.(\ref{fig2})
are similar in nature, while the outward streamlines of $z>0$ planes
explains the pumping shown in Fig.(\ref{fig2}). In addition, as explicit
from the Fig.(\ref{fig1theory}) itself, the axial rotation of the
tracer particles demonstrated in Fig.(\ref{fig6}) and Video 12 of
SI, is also depicted accurately by the closed streamlines on the X-Z
plane.

In Fig.(\ref{fig:2theory}), we plot the flow around two bubbles of
the same size (left panel) and different sizes (middle panel). For
bubbles of the same size, a separatrix is formed and the tracer particles
are attracted to a bubbles depending on its position. On the other
hand for bubbles of different size, since the force acting on the
bubble is proportional to the square of the radius of the bubble,
as can be seen from Eq. (\ref{eq:caseA}) and (\ref{eq:forceM}),
the flow is predominantly towards the larger bubble. This situation
is seen in the second panel of the Fig.(\ref{fig:2theory}). This
behaviour is consistent with the experimental observation demonstrated
in Fig.(\ref{fig3}) and Videos 6-8 in the SI. The last panel shows
the three-dimensional flow around two bubbles. 

For the second case, where two bubbles are generated on the same linear
pattern, we speculate that there is a spontaneous symmetry breaking
in the plane of the wall due to the presence of a temperature gradient
along the linear pattern joining the two bubbles. It should be noted
that the temperature distribution now also has a dependence in the
plane of the wall, hence additional boundary conditions have to be
imposed in Eq.(\ref{eq:gradTemp}), This leads to a modified temperature
distribution around the two-bubble system a with non-vanishing torque
in Eq.(\ref{eq:torqueM}), which finally gives rise to anti-symmetric
contributions leading to in-plane vorticity in the flow, demonstrated
in the Fig.(\ref{fig:3theory}), which is also consistent with the
experimental results of Fig.(\ref{fig4}) and Videos 8, 9, 10, and 11
in the SI. The directions of the flow depend on the following:\\
1. Explicit symmetry breaking which happens when there are bubbles of very different sizes, so that the azimuthal symmetry of the flow is broken in a particular definite pattern due to the presence of a unique temperature gradient determined by the bubble sizes. In these cases, we typically observe counter-clockwise rotation around the bigger bubble.\\
2. The really interesting and intriguing scenario is when we have two similar-sized bubbles, and it is our subsequent observation of clockwise and anticlockwise flows around the bubbles which we attribute to spontaneous symmetry breaking. While we do not have a definitive explanation for the spontaneous symmetry breaking, we feel that geometric factors determine the nature of the flows around the bubbles when they are of the same size. Indeed, this is a situation where both clockwise and anticlockwise flows around each bubble are equally probable (Fig.~\ref{fig4} shows a particular flow configuration), and the system most likely collapses to one of the two possible states (a state being termed as a particular direction of flow around a particular bubble) depending crucially on initial conditions that include: a) which bubble is grown first, b) geometric differences in the shapes of the bubbles, i.e. deviations from a spheroid, c) imperfections in the flatness of the substrate which may provide a certain inclination to the plane of the bubbles, etc. Note that it is quite non-trivial to exactly model these effects, and even challenging experimentally to grow two bubbles of the same size adjacent to each other due to laser intensity fluctuations and the gradual increase in the size of the bubbles due to the self-assembly of particles at their base which prevent the dissipation of heat from the hot-spot into the neighbouring fluid. We are presently studying these effects in detail and plan to report this in future work.

To summarize, the heating of the plane wall by laser beams cause a
temperature distribution in the system. We call this temperature distribution
$\theta^{{\scriptscriptstyle \infty}}$. The distribution of this
imposed temperature field determines the temperature field, $\theta(\mathbf{r})$,
on the surface of bubbles. The surface gradients of $\theta(\mathbf{r})$
manifest itself as the active Marangoni stress on the surface of the
bubble, which drives the exterior fluid flow $\boldsymbol{v}(\mathbf{r})$.
The plane wall, where the bubbles are formed, plays a crucial role
in modifying the exterior fluid flow, which leads to the entrainment
of the mesoscopic particles in the flow.

\section*{Conclusions}

In conclusion, we provide a glimpse of the fascinating possibilities
of manipulating the trajectories of micro-particles by specially engineered
flows driven by micro-bubbles induced by thermo-optical tweezers.
We perform experiments with single and pairs of bubbles in an aqueous
medium, and control microscopic flows in water by creating different
temperature profiles in the vicinity of the micro-bubbles. The bubbles
are nucleated on a glass surface (cover slip or microscope slide)
coated with linear micro-patterns of an absorbing material on which
the tweezers laser beam is focused so as to create a hot spot at the
focal region. The non-uniform temperature across the bubble surface
leads to a surface tension gradient which creates active Marangoni
stress that drives flows in the surrounding liquid. We use polystyrene
microspheres as tracer particles to visualize flow lines. With a single
bubble, we observe self assembly of particles on the bubble surface
as a consequence of flow lines tending to converge along an equatorial
circle of the bubble. Thus, we assemble different-sized polystyrene
particles, as well as magnetic and metallic microparticles, thus establishing
this method as an alternative to conventional optical trapping. This
should be especially useful for confinement of particles having large
scattering cross-sections which renders them difficult to trap using
light forces. The single bubble experiments performed with an asymmetric
temperature gradient across the bubble base leads to the generation
of substantially different flow lines along the axial direction, so
that we have simultaneous attraction and repulsion at different z-planes.
Thus, while some particles assemble on the bubble surface - others
are repelled away, suggesting the use of this mechanism as an efficient
sorting tool. We next manipulate the trajectory of the particles by
engineering the flow around a pair of bubbles that are spatially separated
and have independent temperature profiles around their vicinity, so
that particles assemble differentially on the bubbles depending on
their size with the larger bubble attracting a greater number of particles.
This is again an excellent sorting tool, where continuous modulation
of the separatrix that separates the flow lines between each bubble
(easily achieved by by modulating the laser power which changes the
bubble size), can enable the use of this method as a tunable particle
sorter, with particles of a certain size sticking on one of the bubbles,
while others may pass between them. Finally, we are also able to impart
angular velocity to the tracer particles by subjecting them to vortex
flows achieved by growing bubbles on the same partially conducting
pattern, so that the resultant temperature asymmetry in the azimuthal
direction leads to the formation of closed streamlines in the axial
direction. We observe orbiting of the particles radially and axially,
in both clockwise and anticlockwise direction, very similar to that
produced by angular momentum carrying trapping laser beams having
longitudinal and transverse spins, respectively, and opposite topological
charge. Thus, our design offers an alternative mechanism to induce
angular momentum on mesosopic particles - something as yet believed
to be possible only with photons, with the axial rotation - produced by
field configuration also known as 'photonic wheels' being achieved only very recently, and mostly
for evanescent fields. All our experiments are then validated by analytical
treatments of the experimental conditions - where with a suitable
juxtaposition of the Stokes and heat equations solved in the presence
of appropriate boundary conditions - we are able to match qualitatively
the flow profiles obtained experimentally, thus enhancing our understanding
of the problem in hand. We are working on further interesting mechanisms
for exercising control on particle trajectory in fluids using micro-bubbles,
and have some interesting results with modulated bubbles which shall
be disseminated shortly. We believe that engineered flows using active
stresses may well be the new paradigm in particle manipulation in
fluids in the mesoscopic world, and can enable new frontiers of research
in soft matter physics and micromanipulation. 

\section*{Acknowledgements}
The authors would like to acknowledge Dr. Rajesh Singh and Dr. Ronojoy
Adhikari of DAMTP, Centre for Mathematical Sciences, University of
Cambridge, for their help in developing the theoretical formalism, and Dr. Soumyajit Roy, EFAML, IISER Kolkata for providing the SOMs which were patterned. This work was supported by IISER Kolkata, an autonomous teaching and research institute supported by the Ministry of Human Resource Development, Govt. of India. SG acknowledges DST INSPIRE for fellowship.

\bibliographystyle{apsrev4-1}
\bibliography{references.bib}

\end{document}